\documentclass[journal=jpcbfk,manuscript=article]{achemso}

\usepackage[version=3]{mhchem} 

\author{Diego E. Kleiman}
\affiliation[Biophysics]
{Center for Biophysics and Quantitative Biology, University of Illinois at Urbana-Champaign, Urbana, IL, 61801, USA}
\author{Hassan Nadeem}
\affiliation[BioE]
{Department of Bioengineering, University of Illinois at Urbana-Champaign, Urbana, IL, 61801, USA}
\author{Diwakar Shukla}
\email{diwakar@illinois.edu}
\affiliation[Biophysics]
{Center for Biophysics and Quantitative Biology, University of Illinois at Urbana-Champaign, Urbana, IL, 61801, USA}
\alsoaffiliation[BioE]
{Department of Bioengineering, University of Illinois at Urbana-Champaign, Urbana, IL, 61801, USA}
\alsoaffiliation[CHBE]
{Department of Chemical and Biomolecular Engineering, University of Illinois at Urbana-Champaign, Urbana, IL, 61801, USA}
\alsoaffiliation[PlantBio]
{Department of Plant Biology, University of Illinois at Urbana-Champaign, Urbana, IL, 61801, USA}

\title[An \textsf{achemso} demo]
  {Adaptive Sampling Methods for Molecular Dynamics in the Era of Machine Learning}

\abbreviations{IR,NMR,UV}

\begin{document}


\begin{abstract}
Molecular Dynamics (MD) simulations are fundamental computational tools for the study of proteins and their free energy landscapes. However, sampling protein conformational changes through MD simulations is challenging due to the relatively long timescales of these processes. Many enhanced sampling approaches have emerged to tackle this problem, including biased and path-sampling methods. In this perspective, we focus on adaptive sampling algorithms. These techniques differ from other approaches because the thermodynamic ensemble is preserved and the sampling is enhanced solely by restarting MD trajectories at particularly chosen seeds, rather than introducing biasing forces. We begin our treatment with an overview of theoretically transparent methods where we discuss principles and guidelines for adaptive sampling. Then, we present a brief summary of select methods that have been applied to realistic systems in the past. Finally, we discuss recent advances in adaptive sampling methodology powered by machine learning techniques as well as their shortcomings.     
\end{abstract}

\section{Introduction}
Computer simulation studies have been an invaluable tool to study atomic scale phenomena. Experimental observations and theoretical predictions can in theory be validated through these simulations. Molecular dynamics (MD) simulations are a powerful technique which can probe these molecular systems at atomic scales. MD simulations iteratively solve equations of motion which allows the molecular system to evolve over time steps at the order of 1-2 fs and perform sampling to recover statistical ensembles. Although MD simulations offer an unparalleled insight into the atomic world, there are many limitations to this approach.

MD simulations require the interaction potential to be defined in terms of a force field which in turn make the simulation accuracy dependent upon this choice of parameters. Hence the simulations will not offer desired insights in a general sense, rather only for the application for which the force field has been parameterized. In this regard many force fields have been proposed like CHARMM \cite{MacKerell1998}, AMBER\cite{tian2022last}, GROMOS \cite{Oostenbrink2004}, OPLS \cite{Harder2015}, etc. for simulation of biological as well as materials systems. 
 
Another major limitation that MD simulations suffer from is the timescale problem. MD simulations generally employ an integration timestep of 1-2 fs corresponding to the smallest degree of freedom for the molecular system under study. Many processes of practical interest, especially biological processes like protein folding, ligand-binding, etc. are of the order of milliseconds or even higher. For these processes the sampling probability decays with energy (Boltzmann distribution), so high energy or rare transitions pose a challenge. A traditional long MD simulation can also remain stuck in a metastable basin and fail to sample the conformational landscape as desired.  A non-specialized computer provides a computational speed of the order of nanoseconds per day which would require years of computation to reach the millisecond stage. To deal with this bottle-neck many alternate approaches have been investigated. 

Coarse-graining of the system under study has been a popular approach to study such biological processes. In coarse-graining, sets of atoms are collectively represented by "beads" which act as a representative ``pseudo-atom'' that hopes to capture the chemical behavior of the modeled group of atoms, e.g. MARTINI \cite{Souza2021} is a commonly used coarse-grained force field to this end.  This clustering reduces the number of atoms (reducing the number of motion equations to be solved) hence reducing the computational expense as complexity for MD simulations is $\mathcal{O}(NlogN)$ \cite{Allen1989-aa}, for $N$ number of atoms. Another speed up comes from the idea that because finer degrees of freedom have been coarse-grained therefore the integration timestep could now be increased from ~1-2 fs to ~20-40 fs. Although both these speedups are significant, there are some limitations\cite{Alessandri2019} to this approach as well. The outcome of a coarse-grained simulation is heavily dependent upon the choice of coarse-graining scheme, hence there is a possibility of coarse-graining out potentially important degrees of freedom for our system. Another caveat is that the energy surface gets smoothened out due to this coarse-graining effect; therefore, the simulation time does not equal the actual time. In response, calculations of dynamic quantities must be scaled accordingly.

\begin{figure}[ht!]
\centering
\includegraphics[width=1.0 \textwidth]{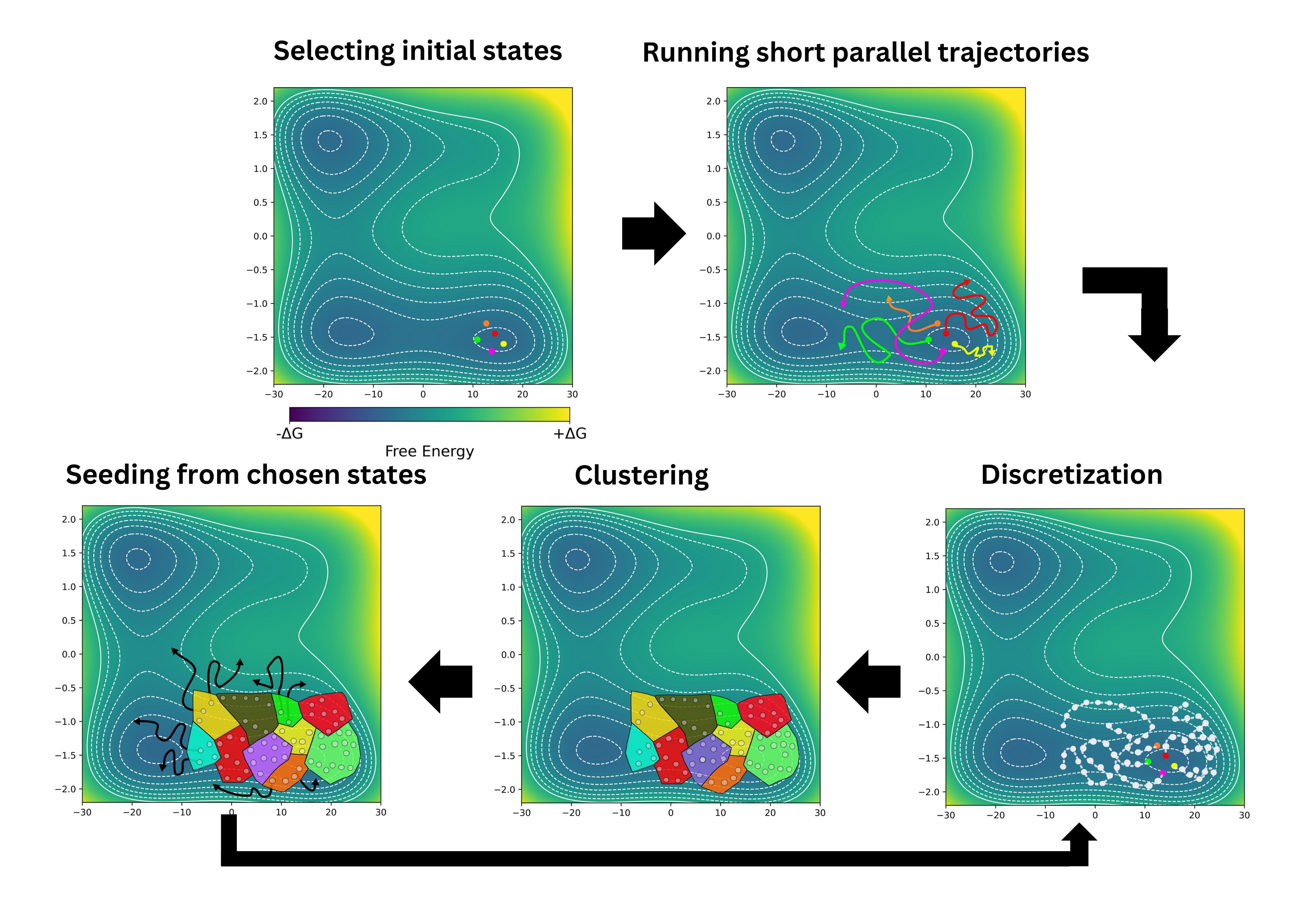}
\caption{Adaptive Sampling, starting from initial states and running short simulations, discretization of conformations, clustering into representative states, and reseeding from states chosen via the chosen scheme.}
\label{fig1}
\end{figure}

Enhanced sampling methods such as adaptive seeding methods, replica exchange methods, localization methods, biasing methods (adaptive and non-adpative) and more, have been proposed as a strong substitute to address this sampling dilemma. The number and types of enhanced sampling methods that have been proposed are too many to mention here and the reader is directed to this excellent review \cite{henin2022enhanced}. Although these methods perform well for specific systems, there are still some drawbacks that limit applicability in a general sense. For example, biased enhanced sampling methods (e.g Metadynamics \cite{Laio2002}) add an external bias to the system, modifying the underlying potential surface, which causes the system to lose kinetic information, while preserving the thermodynamics. Additionally, these methods could potentially sample unphysical conformations due to the external forces. Other techniques (e.g., replica-exchange \cite{Sugita1999}) work well for enthalpic barriers but perform relatively poorly for systems where entropic barriers are dominant.
The class of methods addressed in this paper can generally be described as unbiased adaptive seeding methods. The key idea is to, after an initial run of short trajectories, strategically restart these trajectories based on some criterion. A major portion of adaptive seeding methods are represented by adaptive sampling methods, where, instead of conventional long MD simulations, multiple short simulations are run in parallel and states from the resultant trajectories are selected adaptively to run the next round of simulations. It is the choice of this adaptive selection that distinguishes the different types of methods in this class. The process is illustrated in Figure \ref{fig1}.

The intuition behind this adaptive seeding is to start sampling from states which have been relatively less sampled, and which would be more likely to overcome free-energy barriers in rare events like protein folding. These short trajectories are then ``stitched'' together using Markov State Models, where states are clustered, and probabilities of state transformations are recorded in a transition matrix and analyzed using tICA (time-lagged independent component analysis). The theoretical details have been outlined in the Theory section.
In this perspective we outline the theoretical underpinnings of adaptive seeding as well as major methods that have been developed recently. As of recent, machine-learning inspired adaptive seeding methods have shown promise, which we also touch upon. We also suggest potential future developments in the field as well as point out areas of weaknesses that need to be worked upon.

\section{Theory}

The practice and theory of adaptive seeding methods have developed rather unevenly across their history. For this reason, the theoretical characterization of certain methods is more advanced than for others. Nonetheless, due to the complicated statistical behavior of highly dimensional dynamical systems, simplifying assumptions of varying strength are applied in the derivation of theoretical principles (e.g., expected advantage over naive methods) and principled guidelines (e.g., optimal allocation strategies). The validity of these assumptions is not usually guaranteed in MD simulations. But before exploring these theoretical principles and their underlying assumptions, it is useful to first define the quantities that adaptive seeding methods intend to estimate, as these will motivate the theory.

Adaptive seeding methods are focused on accelerating the sampling of state transitions and the convergence of thermodynamic and kinetic models of the molecular system under study \cite{henin2022enhanced, husic2018markov, zuckerman2017weighted, suarez2021markov}. The free energy landscape encodes the thermodynamics of a molecular system because it provides the probability of observing a conformation under the simulated thermodynamic ensemble, $P(\mathbf{x}) = \frac{1}{\mathcal{Z}}e^\frac{F(\mathbf{x})}{k_BT}$, where $\mathbf{x}$ is a molecular conformation, $\mathcal{Z}$ is the canonical partition function, $k_B$ the Boltzmann constant, $T$ the temperature and $F$ the free energy. 

A kinetic model is a mathematical model that describes the time evolution of a system. In the context of molecular dynamics, a kinetic model is composed of a set of state definitions (generally expressed as boundaries in conformational space) and the average rates of change or mean first passage times (MFPTs) between the states. The MFPT can be defined as the average number of trajectory steps that it takes to reach one state from another. The Hill relation can be used to compute the MFPT from trajectory data,\cite{aristoff2023weighted}

\begin{equation}
    \langle T_B \rangle_A = \left( \frac{d}{dt} \langle N_t \rangle_{\pi}\right)^{-1}
\end{equation}

where $\langle T_B \rangle_A$ is the MFPT from source state A to target state B and $\langle N_t \rangle_{\pi}$ is the number of arrivals on state B at time $t$ given the steady state distribution $\pi$. This estimate of the MFPT is only correct under the assumptions of steady state convergence, which is computationally challenging for complex systems, and recycling boundary conditions (trajectories that reach B are immediately restarted from A). A mathematically rigorous treatment of the Hill relation under molecular dynamics is available in the literature \cite{baudel2023hill}. MFPTs are central to the characterization of molecular systems because, once known, they allow us to calculate other observables through kinetic modeling \cite{noe2008probability}. Depending on the adaptive seeding method utilized, the measured MFPTs might be biased. For example, adaptive sampling results in statistically biased MFPTs, while weighted ensemble takes care of such bias on the fly by assigning weights to trajectories. Methods that produce biased MFPTs employ post-hoc statistical models, like Markov State Models (MSMs) or Generalized Master Equation-based models (GMEMs), to recover the unbiased MFPTs \cite{suarez2021markov, dominic2023building}.

Adaptive seeding can help accelerate the sampling of transition states and kinetic models via two mechanisms, neither of which is exclusive to adaptive seeding methods: trajectory parallelization and selective seeding. Trajectory parallelization refers to running many unbiased trajectories simultaneously. On the other hand, selective seeding is the act of restarting a simulation from a set of specific configurations chosen according to a criterion.  

To analyze the theoretical contribution of each mechanism or, at least, under what circumstances each mechanism is helpful, we can first remove the ``seeding contribution'' from the equation and analyze the advantage of using parallel trajectories only. Interestingly, it is an unrelated field that sheds light over this question. In stochastic resetting the main premise is that a diffusion process (the molecular dynamics simulation) is set back to its original position (conformation) after some random number of time steps \cite{evans2011diffusion}. Clearly, there is no selective seeding under stochastic resetting because the system is always set back to the same seed. We note that in stochastic resetting there is no actual parallelization of trajectories either. However, this is an implementation detail. If they were run in parallel, one could sample permutations from the set (akin to bootstrapping) to recover ``ordered'' trajectories and the statistical analysis holds.

Under stochastic resetting we must consider two random variables: $T$ and $R$\cite{pal2017first}. $T$ is the number of timesteps that it takes an individual trajectory started in state A to reach the target state B without being restarted. $R$ is the maximum possible length of the individual trajectory (after $R$ timesteps, it is restarted). If $T < R$, the FPT under resetting ($T_r$) is measured as $T$. However, if $R \leq T$, $R$ is added to $T_r$ and a new trajectory is sampled. Therefore, we can express $T_r$ in a recursive fashion, $T_r = \min{(T, R)} + \mathbf{1}_{R \leq T}T'_r$ where $T'_r$ is i.i.d. to $T_r$ and $\mathbf{1}_{R \leq T}$ is an indicator function that halts the summation once $T < R$\cite{pal2017first}. Note that in the case of the parallel simulations, we would need to ``discard'' any trajectories in the permutation that come after one where $T < R$. Taking the expectation of this expression, we get\cite{pal2017first}
\begin{equation}
    \langle T_r \rangle = \frac {\langle min(T, R) \rangle}{P(T<R)}.
\end{equation}
This is the ``effective'' MFPT measured by the restarted (or short and parallel) trajectories.

Although the output of the expression depends on the specific probability distributions of $T$ and $R$, it tends to be smaller than the unbiased MFPT ($\langle T_B \rangle_A$) because the distribution of the FPT can be heavy-tailed, a feature that is countered by the restarting procedure. In fact, as it has been proven\cite{reuveni2016optimal, pal2017first} and has been empirically explored\cite{blumer2022stochastic}, if the coefficient of variation (standard deviation divided by the mean) of the unbiased FPT is greater than one, then $\langle T_r \rangle < \langle T_B \rangle_A$. This translates into a speedup in the sampling of state transitions. For example, researchers have shown a toy system where the speedup reaches one order of magnitude with a coefficient of variation of 2.9\cite{blumer2022stochastic}. It must be noted that the actual speedup depends on the distribution of the unbiased FPT, not only on the coefficient of variation\cite{starkov2022universal}. A method to recover the unbiased MFPT from the restarted one was proposed \cite{blumer2022stochastic}, but it seems that the theoretical bound for the error remains elusive. Nonetheless, other statistical methods, such as MSMs\cite{suarez2021markov} and GMEMs\cite{dominic2023building}, might prove useful to recover the unbiased MFPT from restarted trajectories.

The takeaway message from this analysis is that, even without any selective seeding, one could have a considerable speedup in state transition sampling and MFPT convergence from parallelization only. Therefore, when testing new adaptive seeding methods, it is important to include a sensible baseline that accounts for the parallelization advantage. For example, one could compare the proposed technique against another parallel method. If one merely claims that an adaptive seeding method improves upon long, continuous MD simulations without explicitly providing the distribution of the unbiased FPT or its coefficient of variation, then it remains ambiguous if the seeding strategy is actually responsible for the speedup.

Now that we have considered the situation where parallelization alone provides a speedup, we will turn to theoretical results pertaining to the selective seeding advantage. We will focus on three specific approaches based on the theoretical clarity that they bring into the discussion, noting that they are not routinely used in practice and their underlying assumptions are demanding in most cases. 

The first one is termed coupled parallel trajectories\cite{shirts2001mathematical}. In this method, $M$ parallel trajectories are started from the same state. After a single trajectory has sampled a state transition, all the trajectories are moved into the new state. It is assumed that a transition can occur between any two states (all states are connected). By making the additional simplifying assumption that the FPTs between states are exponentially distributed, then we can draw explicit results. For this, note that the problem setup can be expressed as a system of differential equations, $\frac{d}{dt}\mathbf{p} = K\mathbf{p}$, where $K$ is a matrix that contains the transition rates and $\mathbf{p}$ contains the state populations. The long time solution to this system is $\mathbf{p} = \sum_i{c_i \mathbf{v}_i e^{\lambda_it}}$, where $\lambda_i$ and $\mathbf{v}_i$ are the $i$th eigenvalue and eigenvector of $K$, and $c_i$'s are constants that depend on boundary conditions. Further assuming Markovianity and absorbing boundary conditions (the simulations are stopped at the target state), we get that the MFPT to the target state is 

\begin{equation}
    \langle T'_B \rangle_A = \sum_i {\frac{c_i \text{v}_{iB}}{M\lambda_i}} = \frac{1}{M}\langle T_B \rangle_A
\end{equation}
where $\text{v}_{iB}$ is the $B$th component of the $i$th eigenvector. A full derivation of the second equality is available in the original study \cite{shirts2001mathematical}. The key result is that, in some simple cases, a greedy strategy where all trajectories are moved to the newly discovered states as soon as they are seen can provide a linear sampling speedup. Of course, realistic molecular systems are generally too complex for the assumptions to hold; they include unconnected states, kinetic traps, and other features that frustrate this type of greedy schemes. Moreover, when comparing against $M$ parallel but uncoupled trajectories the relative speedup is not linear\cite{shirts2001mathematical}.

The next approach that we will consider offers a more sophisticated view, since it considers selective seeding as a means to reduce the variance in kinetic models rather than simply increasing the sampling rates\cite{hinrichs2007calculation}. The idea behind this method is to perform error analysis on the first non-trivial eigenvalue of the Markovian transition probability matrix and then selectively seed new simulations from the state with the largest contribution to its variance. The result is a sampling scheme that improves the resolution of the slowest relaxation process. To achieve this, two key assumptions are made: (1) the transition probabilities converge to normal distributions after enough transitions have been sampled, and (2) the first-order Taylor expansion around the eigenvalue is a good enough approximation of the effect of small perturbations in the transition probability matrix. To express the selection criterion mathematically, it is useful to define certain quantities first. Let $\Bar{\mathbf{k}}_i = K_i$ be the normalized vector containing all transition probabilities for starting state $i$ and let $\mathbf{s}^{\lambda}_i = \nabla_{\mathbf{k}_i} \lambda \rvert_{\Bar{\mathbf{k}}_i}$ be the sensitivity vector which linearly approximates how much the eigenvalue $\lambda$ will vary when changing an element in the probability transition matrix. These values are used to compute $\Bar{\mathbf{q}}_i = \left( \mathbf{s}^{\lambda}_i \right)^\top \left[ \text{diag}(\Bar{\mathbf{k}}_i) - \Bar{\mathbf{k}}_i\Bar{\mathbf{k}}_i^\top \right] \mathbf{s}^{\lambda}_i$. Then, the variance of the eigenvalue is given by $\sigma^2 = \sum_i {\Bar{\mathbf{q}}_i}/({w_i + 1})$ where $w_i$ is a normalization factor for the transition counts from state $i$ \cite{hinrichs2007calculation}. If we add $m$ new samples to state $i$ (assuming the transition probabilities stay constant), then the state that will result in the highest reduction in the variance will be given by $i = \text{argmax}({\Bar{\mathbf{q}}_i}/{(w_i + 1)} - {\Bar{\mathbf{q}}_i}/{(w_i + m + 1)})$ \cite{hinrichs2007calculation}. When this selection criterion is used, the variance of the first non-trivial eigenvalue decays faster than when using other forms of parallel sampling\cite{hinrichs2007calculation}. This work shows that, given a series of approximations, one can reach an elegant, closed-form solution that determines the criterion for adaptive seeding. Follow up works showed that, in practice, other adaptive sampling schemes perform better in terms of error reduction\cite{weber2011characterization}, a sign that the assumptions used in this analysis (e.g., linearity of eigenvalue perturbations) are too stringent.

The last approach that will be described differs from the previous ones because it was formulated to work under weighted ensembles rather than Markov state models\cite{aristoff2023weighted}. Similarly to the approach described before, the goal is to reduce the variance in a metric of importance to the kinetic model. In this case, rather than minimizing the variance of the first non-trivial eigenvalue, the goal is to reduce the variance in the MFPT from the source state to a sink state under recycling boundary conditions (trajectories that enter the sink are immediately restarted from the source). Since weighted ensemble simulations involve stopping unproductive trajectories (“merging”) and allocating productive ones (“splitting”), the idea behind this approach is to perform these actions following optimal coordinates that guarantee that the variance of the MFPT will be reduced. For this reason, two optimal coordinates must be defined, the flux discrepancy function, $h(\mathbf{x})$, and the flux variance function, $v(\mathbf{x})^2$,

\begin{equation}
    h(\mathbf{x}) = \frac{\langle T_B \rangle_\pi - \langle T_B \rangle_\mathbf{x}}{\langle T_B \rangle_A}
\end{equation}

\begin{equation}
    v(\mathbf{x})^2 = \frac{1}{\tau} \text{Var}_\mathbf{x}\left[ \mathbf{1}_{X_T \in B} + h(X_T) \right]
\end{equation}
where $\mathbf{x}$ is a conformation, $\langle T_B \rangle_\pi$ is the MFPT to the target state from the steady state distribution, $\langle T_B \rangle_\mathbf{x}$ is the MFPT from location $\mathbf{x}$, and $\langle T_B \rangle_A$ is the MFPT from the source state A (which, unfortunately, is the value we were trying to estimate in the first place). Furthermore, $\tau$ is a time interval over which the variance is computed, $X_T$ refers to a trajectory given by the Markovian dynamics of the system, and $\mathbf{1}_{X_T \in B}$ is an indicator function that determines whether $\mathbf{x}$ is in state B at each step in the trajectory. In words, $h(\mathbf{x})$ can be interpreted as the normalized kinetic distance between a point in phase space and the steady state distribution, while $v(\mathbf{x})^2$ gives us the expected change in flux into B given a trajectory started at $\mathbf{x}$.
Of course, it is quite contradictory that we need to know the value that we want to estimate, $\langle T_B \rangle_A$, to compute the optimal coordinates. We also require the steady state distribution, $\pi(\mathbf{x})$, which is typically challenging to compute for complex systems. Nonetheless, the authors of the original study propose to estimate the necessary values between simulation rounds by fitting MSMs with the available data and using the estimates to approximate the optimal coordinates \cite{aristoff2023weighted}. These coordinates are then used with a binning strategy to define regions of phase space which must be allocated the same number of trajectories to optimally reduce the variance in the measured MFPT. Namely, given $H$ bins, we must set endpoints $h_0 < h_1 < \ldots < h_H$ such that \cite{aristoff2023weighted}

\begin{equation}
    \int_{h_i \leq h(\mathbf{x}) \leq h_{i+1}}{\pi(\mathbf{x})v(\mathbf{x}) d\mathbf{x}} = \text{constant.}
\end{equation}
In simpler terms, this means that we should weight a position in phase space not only by the density of the steady state distribution, $\pi(\mathbf{x})$, but also by its contribution to the MFPT fluctuation, $v(\mathbf{x})$.

Following  this allocation rule results in a minimization of the variance of the flux (and therefore the variance on the estimated MFPT). In comparison with the same number of ``brute-force'' parallel trajectories, we obtain a maximum ratio of variance reduction given by \cite{aristoff2023weighted}

\begin{equation}
    \frac{\text{Var}(J_{BF})}{\text{Var}(J_{WE})} = \frac{\int{v(\mathbf{x})^2\pi(\mathbf{x})} d\mathbf{x}}{\left( \int{v(\mathbf{x})\pi(\mathbf{x})} d\mathbf{x} \right)^2}
\end{equation}
where $\text{Var}(J_{BF})$ is the variance in the flux measured from parallel trajectories and $\text{Var}(J_{WE})$ is the variance in the flux measured by the optimally allocated trajectories. Although this expression is not transparent for highly dimensional systems, it can be used to compute the optimal advantage attainable for weighted ensemble simulations in analytical potentials. In particular, this result was used to show that in the low temperature limit the advantage with respect to brute-force simulations is exponential in the largest energy barrier \cite{aristoff2023weighted}.

To summarize, in this section we introduced four theoretical approaches to dissect the advantage of adaptive seeding methods. As we moved from simple parallelization to sophisticated selective seeding criteria, we discussed known theoretical results associated with each approach and their underlying assumptions. Notwithstanding the mathematical transparency afforded by these approaches, different methods have shown better performance in practice. In the next section, we will discuss a wide variety of adaptive seeding methods that have also been applied to realistic systems. After this, we discuss recent advances in the field, with a particular focus on machine learning-based methods.

\section{Methods}

\begin{figure}[t]
\centering
\includegraphics[width=0.9 \textwidth]{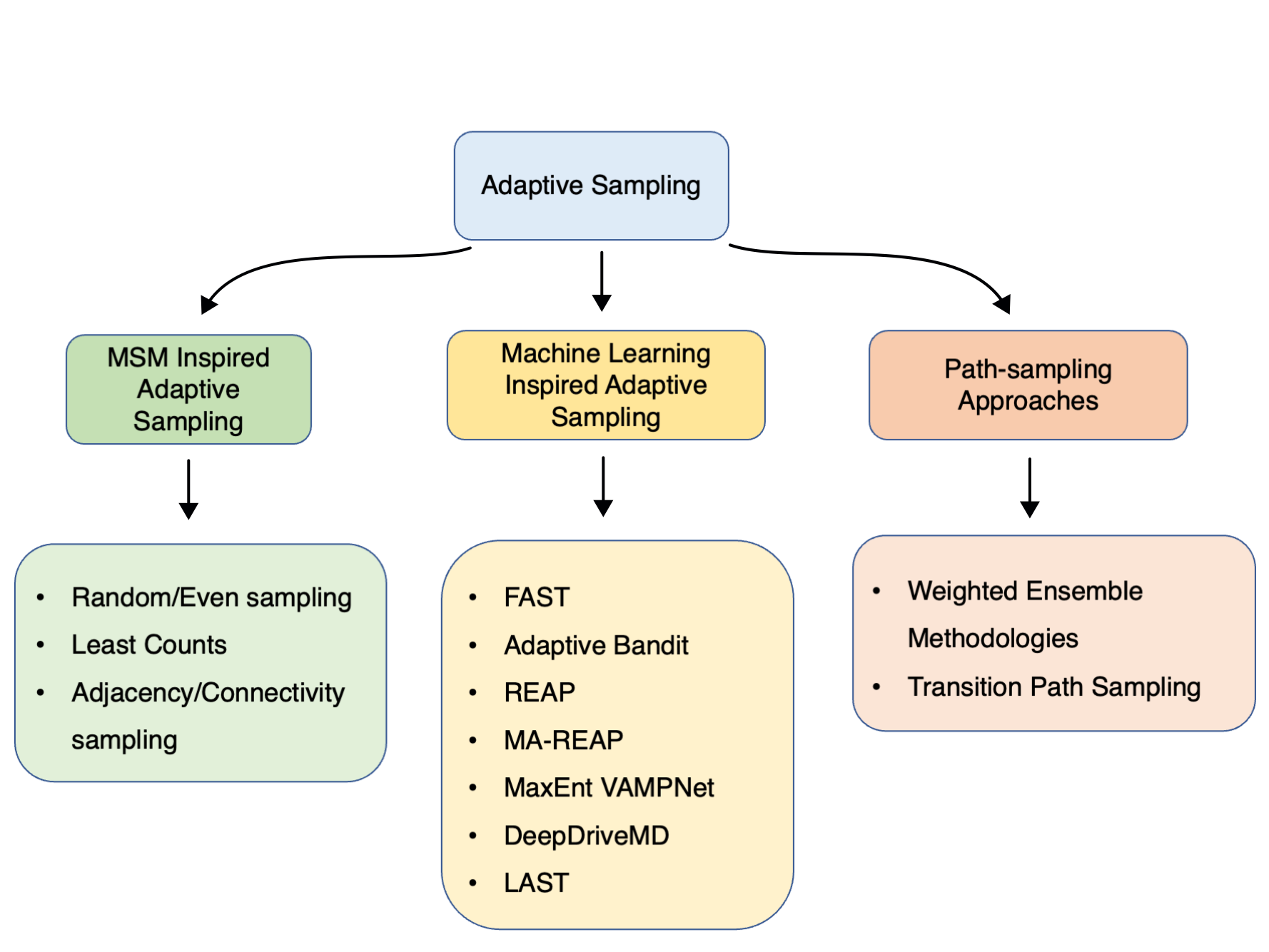}
\caption{\label{fig2}Hierarchy of Adaptive Sampling for enhanced sampling in MD simulations}
\end{figure}

Adaptive sampling (AS) methods can be generally categorized into Markov State Models inspired methods, and Machine Learning inspired schemes. We describe the prominent methodologies in each of these categories followed by a short discussion on other methods, such as path-finding methods, which fall under adaptive seeding. Each of the methods is described briefly to give an overall view of the different adaptive sampling methodologies, readers are referred to relevant literature for an in-depth explanation of respective methods. Figure \ref{fig2} illustrates this broad categorization.

\subsection{MSM-inspired Adaptive Sampling}

\begin{figure}[t!]
\centering
\includegraphics[width=1.0 \textwidth]{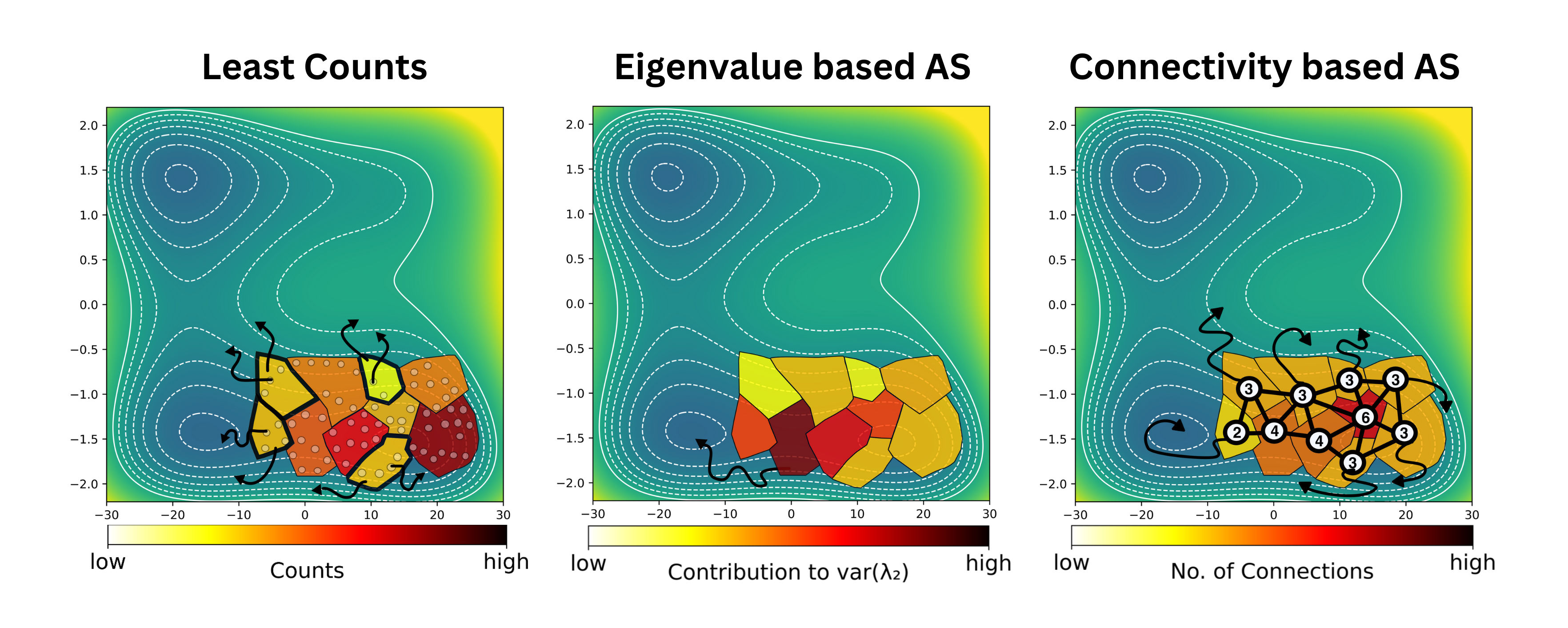}
\caption{\emph{Least Counts}: Least visited states; states with the lowest amount of counts are selected for reseeding. \emph{Eigenvalue based AS}: States contributing most to the first non-trivial eigenvalue of $T_{ij}$ are selected. \emph{Connectivity based AS:} Least connected states; states having the lowest counts in the adjacency matrix are chosen.}
\label{fig3}
\end{figure}

The first class of methods that we describe are Markov state Models inspired adaptive sampling methods. The idea is to initialize simulations to explore and exploit regions of interest using Markov State Models as tools of analysis. This is done by seeding a set of short MD simulations, the resulting states are then clustered according to a kinetic/geometric criterion, and then out of these clusters new states are chosen for seeding according to some methodology. The MSM-inspired machine learning has been extensively employed to investigate biophysical processes such as protein folding\cite{Lane2013}, protein conformational changes\cite{Shukla2014Activation, Kohlhoff2013, Zimmerman2021}, protein-ligand binding\cite{Shukla2019, Chen2021, Dutta2022}, membrane transport\cite{Chan2022,Feng2021,Selvam2019} and protein-protein association\cite{He2021}. It is this choice that distinguishes the method. Most commonly employed MSM inspired AS methods are shown in Figure \ref{fig3}.

One of the first methods was inspired by uncertainty analysis in MSMs \cite{hinrichs2007calculation}. Closed form expressions for distribution of eigenvectors and eigenvalues of the transition matrix were derived. Correspondingly these distributions can be decomposed to calculate contribution of variance to the first non-trivial eigenvalue of the Markovian transition probability matrix $T_{ij}$ for each state and thereby seeding can be selected for states which contribute the maximum to this contribution. The method was shown to significantly increase precision for villin headpiece, but the gain in precision was shown to be linked to the number of states the system had been coarse-grained into. 

A conceptually simpler approach is to do a random \cite{Bowman2010} selection of states to seed from. This method was shown to improve on generalized ensemble methods which fail to overcome entropic barriers at low temperatures, and performed better for sampling of a  hairpin folding system where conformational changes are diffusion controlled.
Variants of these two methods have also been employed. For instance uniform (even) sampling can be done by seeding equally from states. Also, new simulations can be distributed among states in contribution to the uncertainty in the slowest rate.

Count based  sampling \cite{weber2011characterization} is another approach focused on sampling exploration. As the name suggests, states with minimum number of counts are selected, in other words, states which have been less explored in the sampling are preferred for seeding to focus on exploration. 

An adjacency based sampling \cite{weber2011characterization} scheme has also been proposed. The idea is to start new simulations based on a connectivity based criterion. Because MSMs can be approximated generally as being similar to Cayley Trees topologically, the distant states in such a topological structure would be least visited, this is because MSMs can be thought of as network models for transition between states. So states with the least number of connections would be chosen.

\subsection{Machine Learning-inspired Adaptive Sampling}
The next class of methods that we will discuss incorporate Reinforcement Learning ideas to sample the free-energy landscape. This can be done by following a gradient along a known property of interest  and/or using a reward/penalty scheme which penalizes the system moving away from the target state and rewards the converse.

The choice of states to seed from can also be made on a ranking criteria where states are ordered according to a given metric significant for the application at hand. For instance this approach \cite{Doerr2014} has been applied for a trypsin-benzamidine binding system and shown to improve upon traditional high-throughput experiments by an order of magnitude, where the ranking criteria used was mean residence time. A similar novel approach \cite{Shamsi2017} was applied on sampling of pathways for rare conformational transitions, where the choice of states was based upon a distance metric between evolutionarily coupled residues. 

\begin{figure}[ht!]
\centering
\includegraphics[width=1.0 \textwidth]{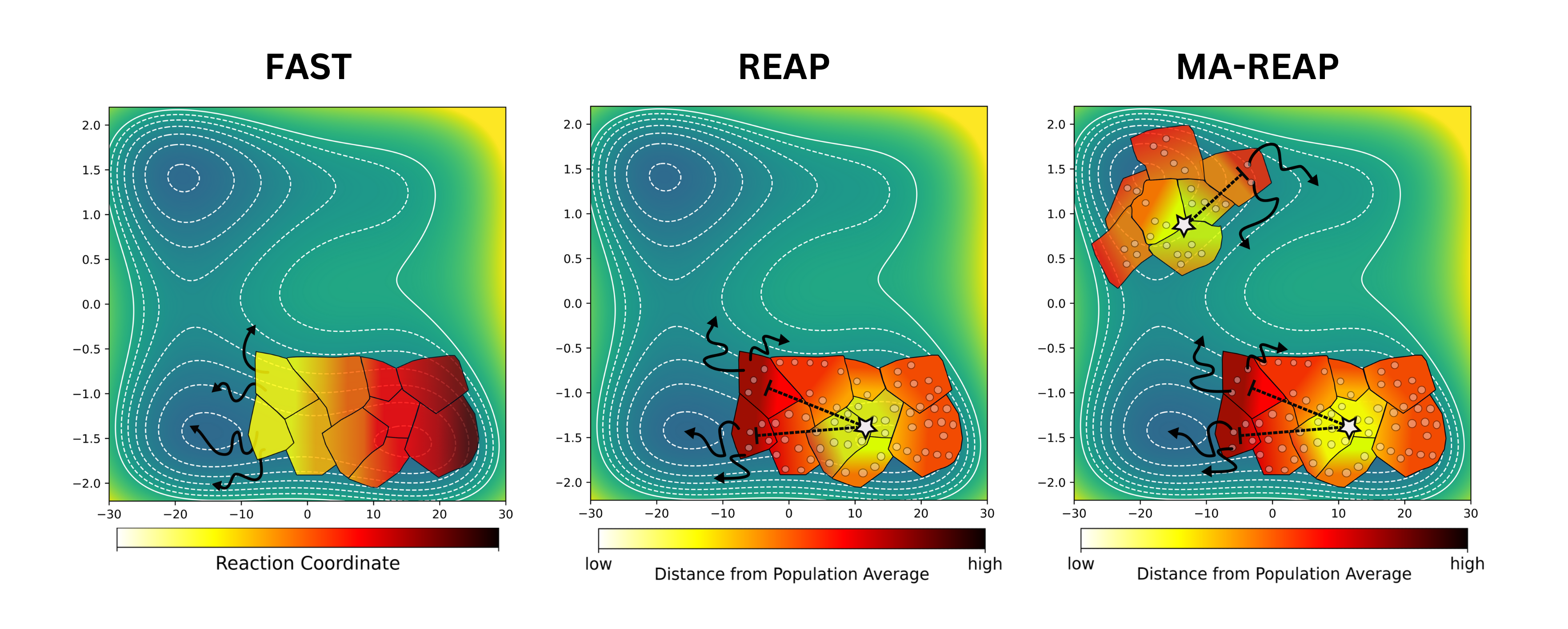}
\caption{\emph{FAST}: States which maximize movement along the gradient of interest are chosen for reseeding. \emph{REAP}: States furthest away from population average (and least visited) are chosen. \emph{MA-REAP}: In multi-agent REAP there are multiple agents that share information and drive the adaptive sampling process after each round.}
\label{fig4}
\end{figure}

Extending this idea of choosing via a ranking scheme, states can also be chosen according to a reward based scheme that favors those states which optimize some property of significance, e.g. RMSD in a protein folding system. Such a scheme named ‘Fluctuation amplification of specific traits’ or ‘FAST’\cite{Zimmerman2015} was proposed and shown to be a significant improvement upon non-directed approaches mentioned above. The test case used was folding of villin protein amongst others. The main idea is that the system of interest will follow an approximately monotonic gradient of some property of interest. For example in a protein folding problem, the transition from an unfolded state to a completely folded state can be thought of as a gradient of solvent accessible surface area which monotonically decreases as the protein folds itself. In essence optimizing along a single reaction coordinate (Figure \ref{fig4}).

Multi-armed bandit problem is a famous problem in combinatorial optimization where an agent faces the choice of a policy that drives an action to return the maximum cumulative reward. AdaptiveBandit\cite{Prez2020} is a sampling algorithm inspired by this problem and aims to maximize the reward, where the reward is set to be the mean of minus free energies of the conformation visited as a result of this action. The inspiration for this reward definition is inspired by the fact that in most MD simulations the aim is to find meta-stable states for example in protein folding problems or ligand binding etc. The policy has to cater for the exploration-exploitation dilemma which is at the heart of this optimization scheme, where the actions have to be exploratory to sample their unknown rewards and also exploitative  to achieve maximum reward from the known best-rewarding space. In the MD perspective this translates to exploring the space of states which is less sampled (exploration) but also focusing on sampling the states which are known to drive the system towards the target end-state (exploitation).

Inspired by this exploration-exploitation dilemma of configuration states in MD simulations, a novel method called REAP\cite{shamsi2018reinforcement} (Reinforcement Learning Based Adaptive Sampling) was proposed. Conceptually, REAP is an extension of the counts method where preference for seeding is given to least visited states, however in REAP this choice is based upon the reward function. This method differs and improves upon the previously described directional methods such as FAST  and AdaptiveBandit (AB). In FAST,  the choice of states follows a gradient along a property of interest (collective variable) and in AB the reward scheme minimizes the free energy of the configurations. In a more generic problem-agnostic framework, these approaches may not always give the best results. For some systems the particular collective variables (CVs, also called reaction coordinates ) or the property of interest may not be known, and instead there may be a set of possibly relevant CVs available. In this scenario a unidirectional gradient approach or a single objective reward strategy will not work. The problem then is to identify relative importance of CVs as well as understanding that this importance (translates to weights in the algorithm) may be changing as the system progresses in the potential energy landscape. REAP solves this problem by dynamically computing these weights so that the system is driven along the CVs which contribute the most to the system moving towards the target state, essentially choosing states which are most ‘distant’ from the average of all conformations in the adaptive sampling round (See Fig 4.). A limitation to this algorithm that is addressed in MA-REAP(addressed next) is that if sampling is started from different states, then pooling the information can result in the rewarding scheme favoring only the states which drive the CVs to have extreme values.

This idea of Reinforcement learning based adaptive sampling has further been extended to an algorithm called MA-REAP\cite{kleiman2022multiagent} (MA for multi-agent). The core of the algorithm resembles REAP closely, the addition has been the idea of having multiple agents that drive the adaptive seeding process after each round. Compared to the single agent traditional REAP algorithm MA-REAP proposes that there may be multiple agents which share data. In REAP, the  action space or the sampled configurations go  through a clustering process, in MA-REAP this is extended in the sense that all agents have stake in each cluster for an action (choice of reseeding) according to the number of configurations present in that cluster because of that action from that agent. In this way, the reward function has a distributed essence according to each agent, allowing each agent to sample along an independent CV and sharing information only when in proximity of other agent’s states. This variation has shown to be an improvement over the traditional REAP and was shown to outperform previously introduced directional methods like FAST and count-based methods as well.

\begin{figure}[ht!]
\centering
\includegraphics[width=0.8 \textwidth]{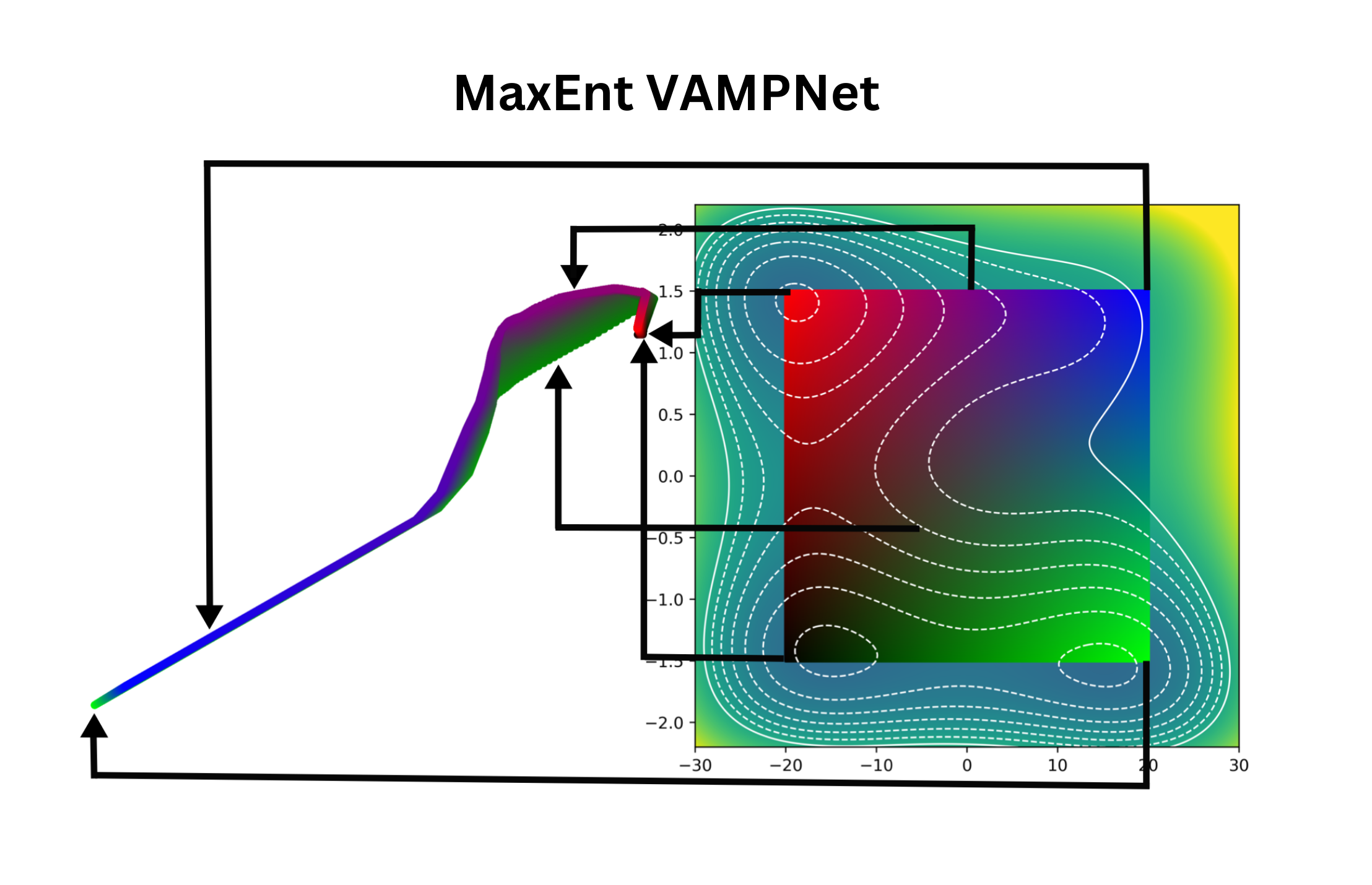}
\caption{VAMPNets are ML models that learn a nonlinear mapping to project the conformational landscape into a new space with kinetically relevant dimensions. Inset on left depicts this latent space projection, where the colors on the internal rectangle drawn on the landscape reveal how this region is mapped by the model. In the technique termed MaxEnt VAMPNet, the inputs that maximize the Shannon entropy of the VAMPNet are chosen as seeds for restarting simulations.}
\label{fig5}
\end{figure}

Another machine learning inspired method to enhance adaptive sampling of biological systems termed ‘MaxEnt VAMPNet’\cite{kleiman2023active} has recently been introduced.  A VAMPNet\cite{mardt2018vampnets} is a deep learning scheme that encodes the entire mapping from molecular coordinates to Markov states, thus reducing the traditional work process of transforming trajectory data into hand-crafted features, dimensionality reduction and estimation of Markov State Models. The idea behind VAMPNets is to learn molecular kinetics using deep neural networks using a variational approach to Markov process, hence ‘VAMP’, Figure \ref{fig5}. MaxEnt VAMPNet method uses the output of VAMPNet as a softmax layer with the interpretation of probability for a microstate to be a kinetically metastable state. The authors then proved that choosing those microstates that maximize an information theoretic measure (Shannon entropy) of the VAMPNet lead to better sampling. A simple summary of the model is to fit VAMPNet to the initial set of trajectories and then choose states with maximum entropy using the output probability of the VAMPNet. The process is repeated by training the model with new trajectories and continuing until desired sampling has been achieved. However, a limitation to the method is that the model cannot be validated at every restart point. A cross-validation step at every iteration would be computationally expensive and an over/under fitting issue would remain undetected. 

\subsection{Other Approaches}
There are various other methodologies that fall under transition path-sampling \cite{Chong2017} approaches for sampling of rare events in biological systems, but we only describe the powerful approach of weighted ensembles for brevity.

Weighted ensemble\cite{Huber1996,zuckerman2017weighted} methods are another class of methods designed to explore pathways for rare states. The basic idea behind weighted ensemble methods in MD simulations is to divide the system into multiple copies, each with a different set of initial conditions. These copies then evolve independently in parallel, and the idea is to replicate simulations that are favoring progression towards the target state while terminating the others. This concept can be thought to  be based on this early idea of a splitting strategy ``When the sampled particle goes from a less important to a more important region, it is split into two independent particles, each one-half the weight of the original''\cite{kahn1951estimation}. So in essence weighted ensemble techniques involve splitting and merging trajectories based on their importance for a sampling criterion. The trajectories are assigned weights, and splitting decreases the weight while merging increases it where the objective is to obtain unbiased observables statistically.

 Random resetting of trajectories, coined ‘stochastic resetting’\cite{blumer2022stochastic} has been recently applied to enhancing the sampling of MD simulations and has shown to result in an increase of an order of magnitude in long time scale processes for simpler systems. In stochastic resetting the set of short simulations is intermittently stopped and restarted where the resetting times are taken with constant steps, termed ‘sharp resetting’ or  can be drawn from an exponential distribution called ‘Poisson resetting’. The idea is that the resetting of the trajectories allows sampling a richer set of pathways, which on average will lead to faster sampling.

\section{Recent Advances}

In this section, we will focus on recent advances in adaptive seeding methodology as well as promising areas of future inquiry. In particular, we will focus on Machine Learning (ML) based techniques. ML has revolutionized many scientific fields, but its impact in the molecular biosciences was catapulted by the high-accuracy of recent protein structure prediction models\cite{jumper2021highly, baek2021accurate, ahdritz2022openfold}. Researchers have been concurrently working on ML models to analyze\cite{mardt2018vampnets, Wehmeyer2018, sultan2018automated, mccarty2017variational} and accelerate\cite{wang2021efficient, Guo2018adaptive} MD simulations. Both sets of tools are becoming increasingly relevant to the field of adaptive seeding simulations. Here, we will discuss how both types of models have been used to improve adaptive seeding MD simulations and how to connect both approaches.

AlphaFold v2 (AF2) \cite{jumper2021highly} and adjacent models\cite{baek2021accurate, ahdritz2022openfold} opened the gates to high-accuracy protein structure prediction from sequence data only. Although structural information in itself is useful to researchers, the output from these models does not provide information on the thermodynamics or kinetics of the protein. For instance, a protein might exist in active and inactive conformational states that interconvert at equilibrium conditions, but AF2 might only predict one of the states. Moreover, we cannot make an inference about the relative free energies of the two states based on which one was predicted by AF2\cite{chakravarty2022alphafold2}. Any information about interconversion rates between the states is also absent from these models. The lack of knowledge about the protein's dynamics can make it impossible to infer its function or mechanism, and therefore complementary methods are required.

MD simulations are excellent tools to computationally resolve the dynamics of proteins, but as discussed in the Introduction, the long-timescale problem turns this approach impractical. As a way to mitigate this issue, researchers have used perturbation-based methods \cite{del2022sampling} to acquire diverse initial conformations for MD simulations from AF2 and similar models \cite{meller2023accelerating, hou2023protein}. Since in general the convergence of kinetic models is highly sensitive to the initial conditions\cite{bhatt2010steady}, adaptive seeding methods stand to gain great speedups simply by improving the prior knowledge of the system. 

For brevity, we will restrict ourselves to two previous studies that applied perturbation-based methods on structure prediction models, but other recent works exist \cite{bodhi2023alphafold}. The first study was used to gather initial structures for parallel simulations of \textit{Plasmodium falciparum} plasmepsin II (PM II) with the intent to sample cryptic binding pockets\cite{meller2023accelerating}. The other method was used to obtain diverse structures of the Shwachman-Bodian-Diamond syndrome protein (SBDS) and the monocarboxylate transporter 1 (MCT1),\cite{hou2023protein} but they were not used for MD simulations although they have the potential to be useful in this regard. 

The first study perturbed the AF2 input by subsampling the multiple-sequence alignment (MSA) of the sequence of interest and enabling stochastic dropout. The process of MSA subsampling consists of restricting the depth of the alignment to a user-selected threshold and randomly selecting the sequence clusters that will be represented. When passed through AF2, the stochastically subsampled MSAs will produce diverse predicted structures that capture some aspects of the dynamics implied by the experimental structures \cite{del2022sampling}. Enabling dropout eliminates a small percentage of the nodes in the neural network during a single forward pass, further perturbing the output of the model. The result of applying this technique to PM II was the prediction of a structural ensemble that partially sampled a known cryptic pocket in the protein. By launching parallel MD simulations from these structures and building a MSM from the trajectories, it was possible to recover the free energy landscape of pocket opening \cite{meller2023accelerating}. These simulations did not require any type of biasing force to find the same free energy basins that could be detected with biased methods \cite{meller2023accelerating}. This study shows that structure prediction models are useful to generate seeds for swarms of unbiased MD simulations. In terms of future directions, it might be interesting to ``close the loop'' and use the simulations to produce new restarting seeds through fine tuning of the structure prediction model, as it has been shown that AF2 can be fine tuned for different specific tasks \cite{motmaen2023peptide, bradley2023structure}.

A different method to perturb the input to structure prediction models consists of splitting the sequence into fragments, predicting the fragments' structure, and connecting them to form a protein. This resembles early approaches to structure prediction, such as Rosetta \cite{rohl2004protein}. A more recent exploration of this approach was presented in MultiSFold \cite{hou2023protein}, a method that combines the variable-length fragment library (VFLib) \cite{feng2022construct} technique with multiple structure prediction models to diversify the predicted conformational ensemble. This method is particularly useful when different prediction models (e.g., AlphaFold and RosettaFold) predict different structures for the same protein. MultiSFold seems to be useful for interpolating structures between functional end points, as it was shown for SBDS and MCT1\cite{hou2023protein}. In the future, it might be interesting to use MultiSFold to find initial seeds for adaptive sampling and then construct new distograms from the simulations. These distograms could then be run through MultiSFold again to obtain new seeds and close the sampling cycle.

Besides using structure prediction models to obtain better initial seeds, ML offers other strengths that have proven useful for adaptive seeding simulations. In particular, the ability to approximate complex non-linear functions from high-dimensional molecular trajectories makes ML models suitable for dimensionality reduction. Finding appropriate low-dimensional projections facilitates many tasks involved in adaptive seeding simulations (e.g., choice of reaction coordinates, clustering, etc.). 

We will focus on three studies that have applied ML models to adaptive sampling, \cite{lee2019deepdrivemd, tian2022last, kleiman2023active} but of course related works exist\cite{ribeiro2018reweighted, ojha2023deepwest, bodhi2023alphafold}. A connection between these three works is that all of them apply a similar simulation-training cycle that consists of launching the simulations, training the model on them, and then using the model to select seeds for new trajectories to restart the loop. Two of these works \cite{lee2019deepdrivemd, tian2022last} use different types of variational autoencoders (VAEs)\cite{kingma2013auto} as their base models, while the third one \cite{kleiman2022multiagent} uses primarily VAMPNets\cite{mardt2018vampnets}. 

Specifically, DeepDriveMD \cite{lee2019deepdrivemd} employs a convolutional VAE (CVAE) while latent space-assisted adaptive sampling (LAST)\cite{tian2022last} employs a VAE parameterized by a feed-forward network with fully-connected layers. Both works use these deep learning (DL) models to learn low-dimensional latent representations that model the probability distribution of the structural ensemble. By selecting the outliers in such a distribution, one can recover the rare conformations that have not been sufficiently sampled, a goal similar to that of count-based adaptive sampling \cite{weber2011characterization}. In LAST, a non-parametric kernel density estimate is applied to obtain the cumulative distribution function in the latent space; the lowest probability samples are then selected as the seeds for the next round of simulation. This is done differently in DeepDriveMD, where the data points are clustered in the latent space with the lowest reconstruction loss using DBSCAN\cite{ester1996density} and then the clusters with fewer than 10 members are selected to restart simulations (with a cap of 150 simulations maximum). The authors of DeepDriveMD also discuss practical aspects, such as distributing the simulations and ML training to different components in high-performance clusters \cite{lee2019deepdrivemd}. LAST was shown to discover the conformational landscape of two proteins (adenilyl cyclase\cite{schlauderer1996structure} and the VIVID flavoprotein\cite{schwerdtfeger2003vivid}) faster than structural dissimilarity sampling \cite{harada2017efficient}, which is another adaptive sampling algorithm. It was estimated to take 40\% of the time to explore the conformational landscape compared to long MD simulations, accounting for the training time of the VAE. Similarly, DeepDriveMD was tested on the Fs peptide\cite{https://doi.org/10.6084/m9.figshare.1030363.v1} and it was found to provide a 2.33 folding speedup compared to parallel simulations.  

MaxEnt VAMPNet\cite{kleiman2023active} is the third approach to be discussed. In this technique, rather than identfying outliers in latent space, a VAMPNet\cite{mardt2018vampnets} is used to classify the discovered conformations into metastable states. Since VAMPNets can assign probabilities, $\mathbf{p}$, of ``belonging'' to a metastable state to each conformation, the Shannon entropy (an information theoretic metric relating a distribution to the uncertainty of the model) is used to score the data points with the formula $H(\mathbf{p}) = - \sum_i{p_i}{\log{p_i}}$. The conformations with the highest entropy are selected for the next round of simulations. The rationale for selecting the structures based on their entropy is that the structures that cannot be clearly placed into a metastable state correspond to transition states or poorly sampled regions of the phase space. When applied to a small peptide (sequence WLALL \cite{scherer2015pyemma}), MaxEnt showed a 3x acceleration in conformational landscape discovery compared to a combination of VAMPNet and count-based adaptive sampling\cite{kleiman2023active} and approximately 60\% higher landscape discovery compared to reinforcement learning-based approaches\cite{shamsi2018reinforcement, kleiman2022multiagent}. Interestingly, the trajectories collected with MaxEnt also produced converged MSMs, while the ones collected with count-based sampling alone did not. When applied to a small protein, the villin headpiece\cite{chiu2005high}, it showed approximately 50\% higher landscape discovery compared to the combination of VAMPNet and count-based sampling.

In summary, this section covered how ML can be used to improve adaptive seeding methods with a particular focus on adaptive sampling. Two avenues to incorporate ML into these workflows were explained: (1) using structure prediction models to augment the prior knowledge of the simulated system and (2) training ML models to rank conformations for optimal seed selection. Combining these two approaches in different ways will probably yield new adaptive seeding methods that surpass the current state of the art. In the next section, we will focus on the challenges that these approaches face.

\section{Challenges}

Incorporating ML models into adaptive seeding algorithms is already showing extremely encouraging results, but this also means that the MD workflows will inherit the challenges associated with these models. With respect to approaches such as DeepDriveMD, LAST, and MaxEnt VAMPNet, the main challenge of using deep models is that fitting them at each simulation round takes considerable computational time and power. Nonetheless, MD simulations continue to be slower, so even when accounting for the training time for ML models, there is a considerable speedup \cite{tian2022last}. Another issue is that these models come accompanied by a host of design choices and optimization variables, although past results demonstrate speedups without an exhaustive hyperparameter tuning \cite{lee2019deepdrivemd, tian2022last, kleiman2022multiagent}. Finally, the data sparsity towards the beginning of the simulation can result in noisy models whose validation scores might fluctuate widely. This could result in noisy selection criteria for simulation restarts. A potential way to prevent this issue would be to train several models initialized randomly and then combine their output using ensemble methods \cite{jeffares2023joint}.

With respect to the use of structure prediction models to seed MD simulations, some limitations can also be noted. Assuming that the quality of the models is good enough to provide accurate conformations, there are issues when the predictions are too similar to a single native structure. When that is the case, launching simulations from several predicted structures might not provide an advantage because the trajectories will be highly correlated. For instance, in the case of MultiSFold\cite{hou2023protein}, if RosettaFold\cite{baek2021accurate} and AF2\cite{jumper2021highly} output similar structures, then the constructed optimization potentials will overlap and the generated ensemble will not be diverse. Similarly, for MSA subsampling of AF2, it might be challenging to gather diverse structures if few experimental structures are known \cite{del2022sampling, stein2022speach_af}. These issues might be tackled by future work through the combination of structure prediction models with MD simulation data sets to produce physically-informed structure prediction of conformational ensembles. 

We must also consider the possibility that the quality of the predicted model is not good enough to confidently employ it as an initial seed in MD simulations. If one starts MD simulations from unphysical conformations of a protein and then attempts to construct a MSM \cite{suarez2021markov}, this could result in disconnected states for which transitions cannot be sampled. One could establish a confidence threshold to accept or reject predicted structures prior to executing MD trajectories by using the pLDDT or similar scores, but a benchmark for this specific purpose is lacking.  

Overall, we have discussed some potential challenges in applying ML models in adaptive seeding algorithms. In particular, we noted that two of the most prominent challenges to the application of ML models in adaptive seeding algorithms are data sparsity and model validation. Potential avenues to mitigate these issues were mentioned.

\section{Conclusions}

In this perspective we have introduced the reader to many aspects of adaptive seeding MD simulations. Theoretically-motivated approaches were presented in the Theory section. In the Methods section, several practiced approaches were summarized. Finally, we introduced recent advances that combine ML methods with adaptive seeding simulations in different ways and discussed their limitations or potential improvements. The authors hope that the ideas presented in this text will inspire further innovation in the area of adaptive sampling for MD simulations. 

\begin{acknowledgement}
The authors acknowledge support from the National Science Foundation Early CAREER Award (NSF MCB-1845606).

\end{acknowledgement}




\providecommand{\latin}[1]{#1}
\providecommand*\mcitethebibliography{\thebibliography}
\csname @ifundefined\endcsname{endmcitethebibliography}
  {\let\endmcitethebibliography\endthebibliography}{}

\end{document}